\documentclass[
,final 
]
{aipproc}
\usepackage[T1]{fontenc}
\usepackage[latin2]{inputenc}
\usepackage{polski}
\usepackage{graphicx}
\usepackage[normalem]{ulem} 
\usepackage{anysize}
\usepackage{fancyhdr}
\usepackage{calc}
\usepackage{geometry}
\usepackage{epsfig}
\usepackage{amssymb}
\usepackage{amsmath}
\usepackage{indentfirst}
\usepackage{chngcntr}
\usepackage{color}

\layoutstyle{8x11single}

\begin{document}

\title{Twist expansion of differential cross-sections of forward Drell-Yan process}

\classification{13.85.Qk, 12.38.Cy}
\keywords {twist expansion, forward Drell-Yan, low x}


\author{Tomasz Stebel}{
address={M. Smoluchowski Institute of Physics, Jagiellonian University,\\
S. Lojasiewicz str.\ 11, 30-348 Krakow, Poland.}
}

\author{Leszek Motyka}{
address={M. Smoluchowski Institute of Physics, Jagiellonian University,\\
S. Lojasiewicz str.\ 11, 30-348 Krakow, Poland.}
}

\author{Mariusz Sadzikowski}{
address={M. Smoluchowski Institute of Physics, Jagiellonian University,\\
S. Lojasiewicz str.\ 11, 30-348 Krakow, Poland.}
}

\begin{abstract}
Forward Drell-Yan process at the LHC is a sensitive tool for investigating higher twist effects in QCD. The expansion of all Drell-Yan structure functions is performed assuming GBW saturation model and the saturation scale plays the role of the hadronic scale of OPE. We show that the Lam-Tung relation is broken at twist 4. The results open the way for a forthcoming analysis of multiple scattering and higher twist effects.
\end{abstract}

\maketitle


\subsection{Introduction}
Large Hadron Collider (LHC) opens new kinematic regions in high energy physics. One of the most interesting of those areas is a region of 
small 
Bjorken-$x$. The most promising process at the LHC for investigating QCD effects below $x \simeq 10^{-6}$ and moderate energy scales is a forward Drell-Yan (DY) 
\cite{lhcb0}. 
Such a 
small $x$ at parton density scale $\mu^2 > 6$~GeV$^2$ is about two orders of magnitude smaller than measurements at HERA. Due to the forward character of the process the LHCb detector is the most suitable for those measurements \cite{lhcb0}. 
The differential DY cross-section may be parametrized in terms of four structure functions 
that describe
hadronic part of the process. One of the main goal of our work is to perform Operator Product Expansion (OPE) for these structure functions. The leading twist, namely 
twist~2,
is well known from theoretical and experimental side, in particular it could be computed using known Parton Density Functions 
(PDFs). However higher twists 
are 
neither well understood theoretically nor explored experimentally yet.
On the other hand, in the small $x$ regime higher twist effects and multiple scattering corrections are important. Understanding of those effects is necessary for example to improve 
the determination precision of
parton densities at the leading twist. Due to possible very 
small~$x$ in the process, analysis of higher twists in Drell-Yan 
should provide new valuable information.

For our purpose the most convenient description of DY process is based on so-called helicity structure functions $W_L, W_T, W_{TT}, W_{LT}$ (see \cite{LamTung1}). In this approach one 
factorizes
leptonic and hadronic degrees of freedom by contracting both 
hadronic and leptonic
tensors with virtual photon polarization vectors 
(PPVs).
The
tensor
reduces to a distribution of lepton angles $\Omega=(\theta,\phi)$ in lepton pair 
center-of-mass
frame while 
the
result of contraction of 
hadronic
tensor with different PPVs are the $W$-structure functions. The inclusive DY cross-section is given by the formula: 
\begin{eqnarray}
\frac{d\sigma}{d x_F dM^2 d \Omega d^2 q_\bot}= \frac{\alpha^2_{em}\sigma_0}{2(2\pi)^4 M^4} \left[ W_L (1-\cos ^2 \theta) + W_T (1+\cos ^2 \theta) + W_{TT}(\sin^2\theta \cos 2\phi)+W_{LT}(\sin2\theta \cos \phi) \right],
\label{sigAsWcomb}
\end{eqnarray}
where $x_F$ is a fraction of projectile's longitudinal momentum taken by virtual photon, $M$ is an invariant mass of leptons pair, $q_\bot$ is transverse momentum of virtual photon and 
the 
constant $\sigma_0$ gives 
the 
dimension. 
The 
form of $W$-structure functions depends on arbitrary choice of axes (which defines PPVs) in lepton pair 
center-of-mass
frame. 
In this paper we use a frame with 
the 
$Z$~axis antiparallel to the target's momentum and 
the 
$Y$ axis orthogonal to
the
reaction plane (in \cite{LamTung1} this frame is called $t$-channel helicity frame).

\subsection{Inclusive cross section for forward Drell-Yan}
\begin{figure}
\includegraphics[height=4 cm]{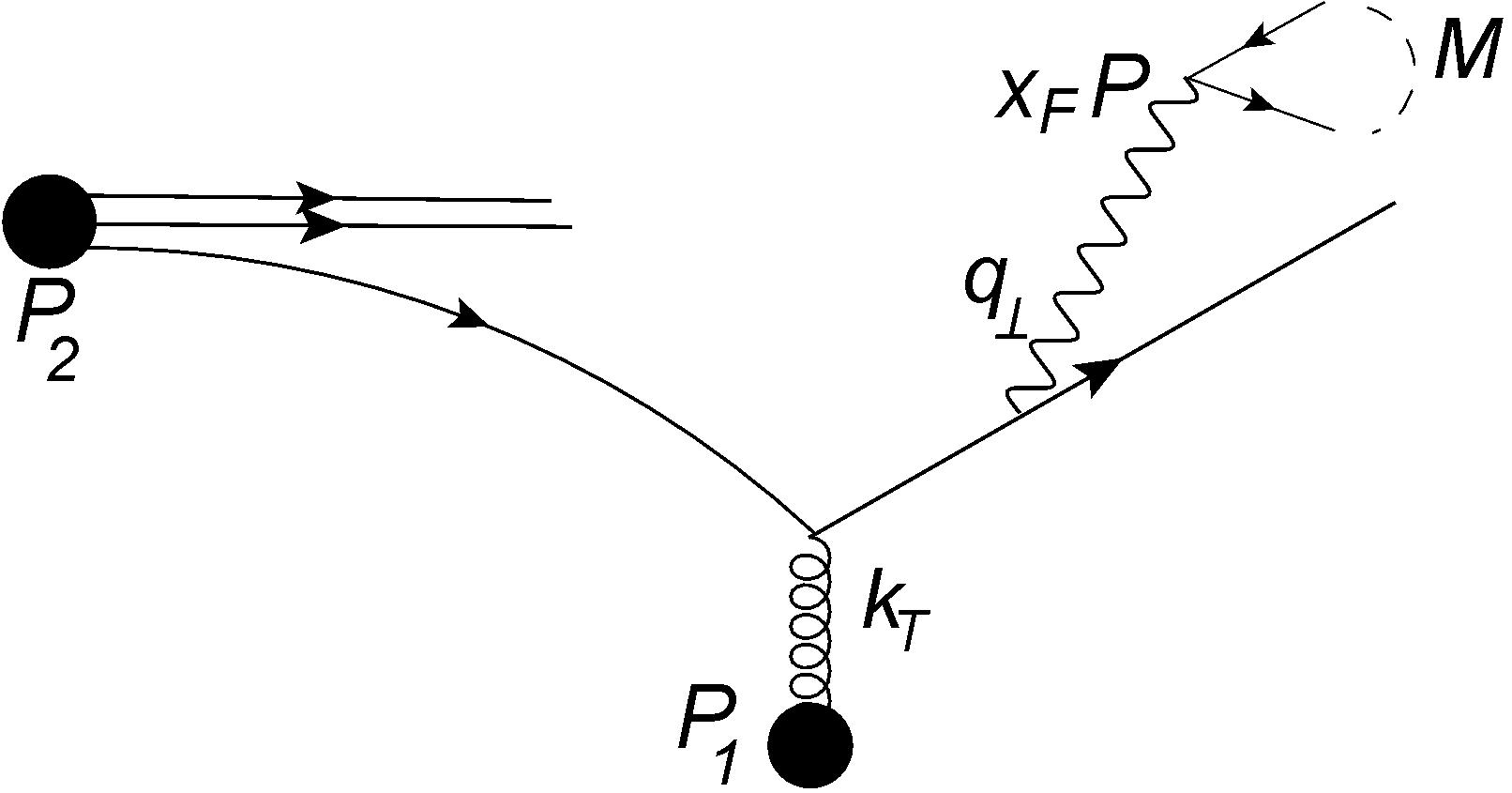}
\includegraphics[height=4.5 cm]{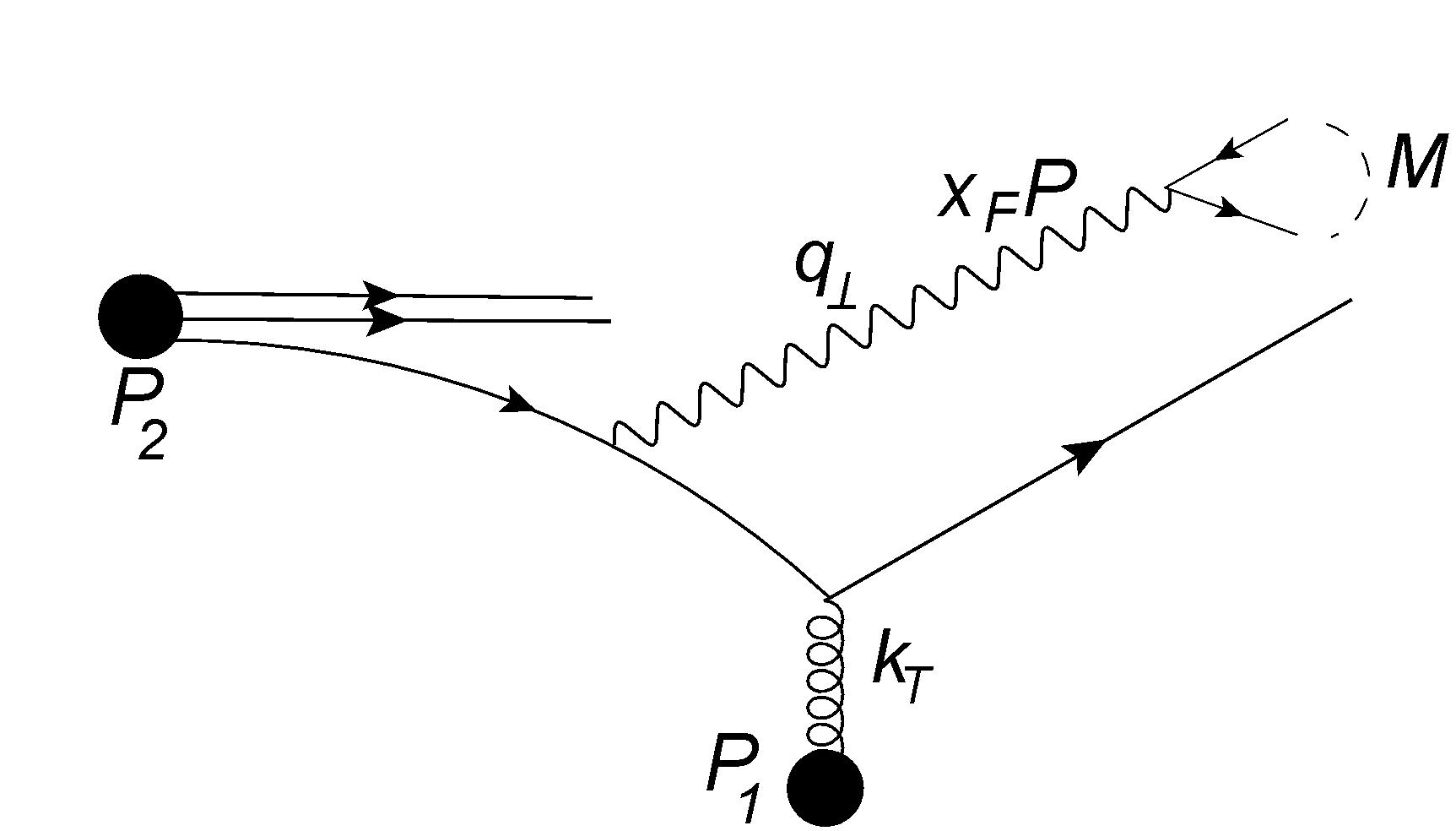}
\caption{Dominant diagrams for forward Drell-Yan process in target rest frame. Photon carries $x_F$ part of projectile's momentum. $M^2$ is photon's virtuality and the square of invariant mass of lepton-antilepton pair. Transverse momenta of photon and gluon is denoted as $q_\bot$ and $k_T$, respectively.}
\label{leadingDiagrams}
\end{figure}

In forward Drell-Yan scattering at the LHC there is a strong asymmetry in 
longitudinal momentum fractions, $x$, of the colliding partons. 
We denote 
the 
proton from which comes
the
fast parton ($x_2\sim 0.1$) by $P_2$ and call it 
the
projectile while 
the
proton with 
the
low-momentum parton 
($x_1 <10^{-5}$) by $P_1$ and call it 
the
target. 
For very small~$x$
the
gluon density is much 
larger
than 
other PDFs 
so 
the dominant diagrams are of the form given in Fig.\ \ref{leadingDiagrams}. We draw these diagrams such that 
the
target is at rest --- it is 
the
most suitable frame for our calculation. Since 
the
energy in 
colliding hadrons' center-of-mass frame, 
$E$,
is much larger than other scales (such as $M$ or $q_\bot$)
we drop in calculations non-leading terms in $1/E$ expansion.

The diagrams from Fig. \ref{leadingDiagrams} may be calculated using $k_T$-factorization framework, 
quite effective in high energy limit of QCD.
Using $k_T$-factorization the inclusive cross section is given as:
\begin{eqnarray}
\frac{d\sigma}{d x_F dM^2 d \Omega d^2 q_\bot}= \frac{\alpha_{em}}{(2\pi)^2(P_1\cdot P_2)^2 \ M^2} \int_{x_F}^1 dz \frac{1}{1-z} \frac{\wp(x_F/z)}{x_F^2} 
\int d^2 k_T \frac{2\pi \alpha_s}{3} \frac{f(\bar{x}_g,k_T^2)}{k_T^4} L^{\tau \tau'}(\Omega) \tilde{\Phi}_{\tau \tau'} (q_\bot,k_T,z)
\label{dsigma}
\end{eqnarray}
and it consists of the following parts:
\begin{itemize}
\item $\wp(x_F/z)$ is a collinear parton distribution function for the 
projectile.
\item $f(\bar{x}_g,k_T^2)$ is an unintegrated gluon density describing interaction of fast quark with proton $P_1$.
\item $ L^{\tau \tau'}(\Omega)$ is a lepton tensor contracted with PPV which reduces to angular coefficients like in (\ref{sigAsWcomb}).
\item Impact factor
$\tilde{\Phi}_{\tau \tau'} (q_\bot,k_T,z)= \sum_{\lambda_1,\lambda_2=\pm} \ A_{\lambda_1,\lambda_2}^{(\tau)}(\vec{q}_{\bot})^\dagger A_{\lambda_1,\lambda_2}^{(\tau')}(\vec{q}_{\bot})$, where
$A_{\lambda_1,\lambda_2}^{(\tau)}(\vec{q}_{\bot})$ is a hard part of amplitude describing emission of virtual photon with polarization $\tau$. Indices $\lambda_1,\lambda_2$ are the helicities of incoming and outgoing quarks. 
\end{itemize}

In description of high energy scattering it is convenient to use the color dipole model in which the unitegrated gluon density in (\ref{dsigma}) is replaced by an
 (equivalent in the leading logaritmic approximation) color dipole cross-section 
\cite{NZ}. This approach was proven to be successful in description Deep Inelastic Scattering (DIS) 
and diffractive DIS 
data from HERA (see \cite{GolecWusthoff}).
In the application of the color dipole model to the forward DY scattering we
follow Refs.\ \cite{Brodsky,Kopeliovich}.
We shall use dipole cross section 
$\hat{\sigma}$ 
fitted to the DIS data in
our DY calculations. 

The dipole cross-section is related to 
the
unintegrated gluon density $f(\bar{x}_g,k_T^2)$ as follows:
\begin{equation}
\hat{\sigma}(r)=\frac{2\pi \alpha_s}{3} \int d^2 k_T \frac{ f(\bar{x}_g,k_T^2) }{k_T^4} \ \big|1- e^{-i \vec{k}_T\cdot \vec{r}} \big|^2.
\end{equation}

Inverting this formula one can rewrite ($\ref{dsigma}$) in 
transverse
position space. Then the helicity structure functions (\ref{sigAsWcomb}) are easily related to the impact factors:
\begin{equation}
W_{i}=\frac{2(2\pi)^4 M^4}{\alpha^2_{em}\sigma_0} \int_{x_F}^1 dz \ \wp(x_F/z)
\int d^2 r \ \hat{\sigma}(r) \Phi_{i} (q_\bot,r,z), \ \ \ \ \ \ \textrm{for} \ \ i=\left\{ L,T,TT,LT \right \}.
\label{Wphi}
\end{equation}
Here we introduced impact factors with definite helicity $\Phi_{L},\ldots,\Phi_{LT}$ which are linear combinations of $\tilde{\Phi}_{\tau \tau'}$'s Fourier-transformed to position space.

In order to find 
the
twist expansion of (\ref{Wphi}) 
we follow methods developed in Refs.\ \cite{twist1,twist2,GolecLew} and 
apply the Mellin transformation to 
the
last integral:
\begin{equation}
W_{i}=\int_{x_F}^1 dz \ \wp(x_F/z)
\int_C \frac{ds}{2\pi i} \ \left ( \frac{z^2 Q_0^2}{M^2 (1-z)} \right) ^s \tilde{\sigma} (-s) \hat{\Phi}_{i} (q_\bot,s,z),
\label{Wphihat}
\end{equation}
where $\tilde{\sigma}(-s)$ and $\hat{\Phi}_{i} (q_\bot,s,z)$ are Mellin transforms of $ \hat{\sigma}(r)$ and $\Phi_{i} (q_\bot,r,z)$, respectively. 
The contour $C$
is a straight vertical line 
in the 
complex~$s$ plane 
which should be closed from the positive side.

Note that in ($\ref{Wphihat}$) we have explicitly three energy scales: one soft $Q_0$ which is a saturation scale (coming from dipole cross section $\hat{\sigma}$) and two semihard scales: $M$ and $q_\bot$. OPE is here given in terms of positive powers of 
the
soft scale $Q_0$.

\subsection{Twist expansion and Lam-Tung relation}

In order to perform twist expansion one should choose 
a
model of 
the
dipole cross-section $\hat{\sigma}$. We adopt 
the
Golec-Biernat and W\"{u}sthoff model \cite{GolecWusthoff}
\begin{equation}
\hat{\sigma}(\vec{r})=\sigma_0(1-e^{-r^2 Q_0^2/4}),
\end{equation}
where $Q_0$ is a 
$x_1$-dependent
saturation scale.

Mellin transform of such function is particularly simple: $\tilde{\sigma}(-s)=-\sigma_0 \Gamma(-s)$. Since $\hat{\Phi}_{i} (q_\bot,s,z)$ are analytic functions of $s$ in positive half-plane, the poles of integrand in (\ref{Wphihat}) come from $ \Gamma(-s)$ function. These are positive integers so 
the
integral is a sum of infinite number of residues which are proportional to $Q_0^{2k}$ with $k=1,2,\ldots$.

As an example
of a result, for $W_L$ we get twist~2 of the form:
\begin{equation}
W_L^{(2)}= \frac{Q_0^2}{M^2} \int_{x_F}^1 dz \ \wp(x_F/z) \frac{4M^{6}\ q_\bot^2 (1-z)^2}{\left[q_\bot^2+M^2(1-z) \right]^4},
\label{WLtw2}
\end{equation}
and twist 4:
\begin{eqnarray}
W_L^{(4)}=
\frac{Q_0^4}{M^4} \int_{x_F}^1 dz \ \wp(x_F/z) z^2 \ \frac{4 M^{8} \left[7 q_\bot^2 -10 M^2 q_\bot^2 (1-z) +M^4 (1-z)^2 \right] (1-z)^2}{\left[q_\bot^2+M^2 (1-z) \right]^6}.
\label{sigmaLtw4}
\end{eqnarray}

For experimental 
searches
of higher twists the most interesting 
are quantities that vanish at the leading twist. 
In 
the
DY process such a quantity 
may
constructed using 
the
Lam-Tung relation \cite{LamTung2}. This relation was proven for the parton model and in terms of helicity structure functions takes
the
form \cite{Gelis}:
\begin{equation}
W^{\mathrm{par}}_L-2 W^{\mathrm{par}}_{TT}=0.
\label{relLT}
\end{equation}

At twist~2 
$W_{TT}$ is given by:
\begin{equation}
W_{TT}^{(2)}= \frac{Q_0^2}{M^2} \int_{x_F}^1 dz \ \wp(x_F/z) \frac{2M^{6}\ q_\bot^2 (1-z)^2}{\left[q_\bot^2+M^2(1-z) \right]^4},
\label{WTTtw2}
\end{equation}
and comparing it with (\ref{WLtw2}) we immediately see that
the
Lam-Tung relation (\ref{relLT}) is satisfied at the leading twist, as expected.

At the next-to-leading twist, namely twist 4, relation is broken:
\begin{equation}
W^{(4)}_L-2 W^{(4)}_{TT}= \frac{Q_0^4}{M^4} \ \int_{x_F}^1 dz \ \wp(x_F/z) z^2 \frac{4M^{8} (1-z)^2}{\left[q_\bot^2+M^2 (1-z) \right]^4}.
\label{LTrelTw4}
\end{equation}

It is 
a
well known fact that (\ref{relLT}) is violated also by 
higher order
QCD corrections, however at the very small~$x$ 
the
contribution coming from higher twists is sizeable comparing to them.

\subsection{Twist expansion of $q_\bot$-inclusive cross section}
In \cite{GolecLew} 
the
twist expansion for forward DY process was performed for cross-section inclusive in $q_\bot$ and lepton angles ($\theta$, $\phi$), then only $W_L$ and $W_T$ give 
the
contribution. To fill 
this
gap we performed
the
twist expansion for $W_{TT}$ and $W_{LT}$ integrated over $q_\bot$. This is interesting also since angular distribution inclusive in $q_\bot$ is easier to measure. 

We 
define
$\tilde{W}_i=(2\pi M^2)^{-1} \int W_i \ d^2 q_\bot$ which for 
the 
cross-section $\frac{d\sigma}{d x_F dM^2 d \Omega }$ are analogues of $W_i$. At first glance
the 
twist expansion could be performed in the same way as 
in the 
differential 
case, however taking residue for $s=1$ for $\tilde{W}_T$ we get:
\begin{equation}
\frac{Q_0^2}{M^2} \int_{x_F}^1 dz \ \wp(x_F/z) \frac{1+(1-z)^2}{1-z} \frac{\pi M^2 }{3}.
\end{equation}
This is clearly 
a
divergent integral over $z$. This means that 
the
integration over $ q_\bot$ introduces double poles in expressions for $\tilde{W}_i$. We shall follow \cite{GolecLew} to obtain convergent expressions 
in the twist expansion.
Then for example: 
\begin{equation}
\tilde{W}^{(2)}_{T}=\frac{Q_0^2}{4 M^2}\Bigg\{ \wp(x_F) \left[-1+\frac{4}{3}\gamma_E -\frac{2}{3} \ln\left( \frac{Q_0^2}{4 M^2(1-x_F)} \right) + \frac{2}{3} \psi (5/2) \right]+\frac{2}{3} \int_{x_F}^1 dz \ \frac{\wp(x_F/z) [1+(1-z)^2] - \wp(x_F)}{1-z}.
\Bigg\}
\end{equation}
Due to additional poles coming from integral over $z$ we get also nonzero odd twists --- in our frame only for $\tilde{W}_{LT}$:
\begin{eqnarray}
\tilde{W}^{(3)}_{LT}= const \ \frac{Q_0^3}{M^3} \wp(x_F),
\end{eqnarray}
where $const \approx 0.593$.

\section{Conclusions and Outlook} 

The forward Drell-Yan scattering
is a 
promising process for searching of higher twists because LHC is expected to provide data for kinematic region of very small~$x$ where they are sizeable. 
The 
quantity $W_L-2 W_{TT}$ 
should be particularly
useful for such 
searches
since it 
vanishes
at the leading twist. 
In this talk we briefly presented theoretical calculations of the
twist expansion of the forward Drell-Yan cross-section. The full results and predictions of higher twist contributions for the LHC will be given in a
forthcoming paper \cite{MSS}. We emphasize that precise measurements of angular distribution of the forward Drell-Yan could be 
essential for understanding higher twists effects in QCD.

\begin{theacknowledgments}
Authors would like to thank the organizers of the International Workshop on Diffraction in High-Energy Physics 2014, Primosten, Croatia, for the interesting meeting. 
Support of the Polish National Science Centre grant no.\ DEC-2011/01/B/ST2/03643 is gratefully acknowledged. 
TS acknowledges support in scholarship of Marian Smoluchowski Scientic Consortium Matter Energy Future from KNOW funding. 
\end{theacknowledgments}

\bibliographystyle{aipproc} 

\end{document}